\documentclass[a4paper,11pt]{article}
\usepackage{pos}
\usepackage{bm}
\usepackage{times}
\usepackage{braket}
\usepackage{hyperref}
\usepackage{subfigure}
\usepackage{graphicx}
\usepackage{slashed}
\usepackage{orcidlink}
\usepackage{multirow}
\usepackage{amsmath}
\usepackage{mathtools}

\definecolor{green}{rgb}{0,0.6,0}

\newcommand{\lhc}{{\rm lhc}}
\newcommand{\rhc}{{\rm rhc}}

\newcommand{\be}{\begin{equation}} 
\newcommand{\ee}{\end{equation}}
\newcommand{\bea}{\begin{eqnarray}} 
\newcommand{\eea}{\end{eqnarray}}
\newcommand{\beas}{\begin{eqnarray*}} 
\newcommand{\eeas}{\end{eqnarray*}}

\title{Left-hand cut problem in lattice QCD and an EFT-based solution}
\author*[a]{Lu Meng}
\author[a]{Vadim Baru}
\author[a]{Evgeny Epelbaum}
\author[a]{Arseniy~A.~Filin}
\author[a]{Ashot~M.~Gasparyan}

\affiliation[a]{Institut f\"ur Theoretische Physik II, Ruhr-Universit\"at Bochum, D-44780 Bochum, Germany }

\emailAdd{lu.meng@rub.de}
\abstract{Lattice QCD has become an essential tool for studying the hadron-hadron interaction from the first principles. However, when extracting infinite-volume scattering parameters from finite-volume energy levels, the traditional Lüscher formula encounters limitations due to the left-hand cut induced by long-range interactions such as the one-pion exchange. In this work, we propose an alternative approach based on  chiral effective field theory combined with a Hamiltonian method in the plane wave basis. By solving a Schrödinger-like equation in the finite volume, our method connects the finite-volume energy spectrum with infinite-volume observables, while systematically incorporating the long-range physics and solving the left-hand cut problem. The use of the plane wave basis mitigates issues related to partial wave mixing. Our numerical results for $DD^*$ scattering at $m_\pi \approx$ 280 MeV demonstrate that this approach overcomes the limitations of the Lüscher method and points towards a resonance interpretation of the $T_{cc}(3875)$ state, as opposed to the virtual state predicted by traditional analyses.}

\FullConference{The 11th International Workshop on Chiral Dynamics (CD2024)\\
 26-30 August 2024\\
Ruhr University Bochum, Germany\\}

\begin{document}
\maketitle

\section{Introduction}

\begin{figure}[h]
    \centering
    \includegraphics[width=0.8\linewidth]{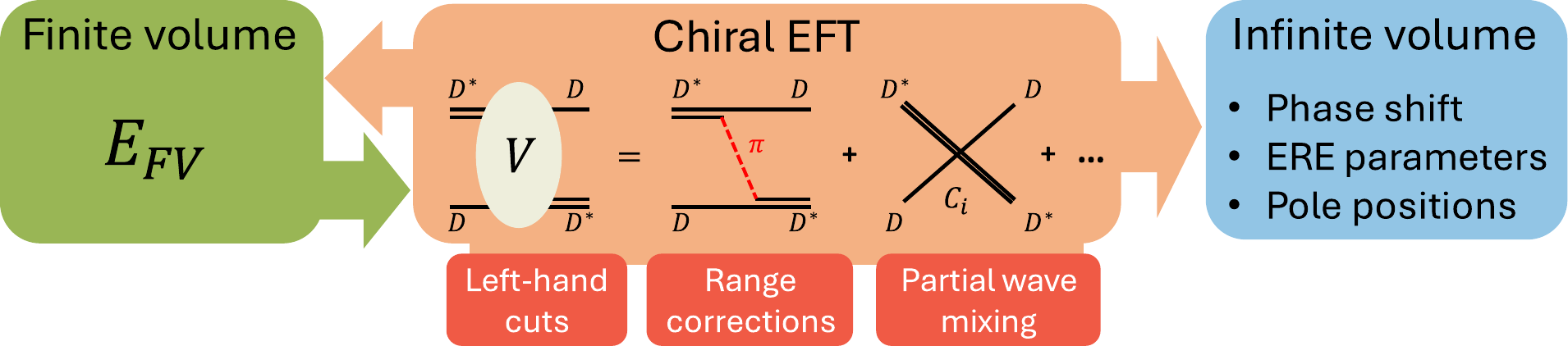}
    \caption{Schematic illustration of the approach used in this study: $ V $ represents the effective potential in chiral EFT, incorporating the one-pion exchange  and contact interactions. $ E_{FV} $ denotes the finite-volume energy levels from lattice simulations, which serve as input.}
    \label{fig:illu}
\end{figure}

Currently, lattice QCD is perhaps the only method to calculate hadron-hadron interactions from first principles. This approach involves implementing QCD on a discretized space-time lattice and computing observables via finite volume (FV) Monte Carlo simulations.  The raw data from lattice QCD simulations are the FV energy levels. Traditionally, Lüscher’s formula~\cite{Luscher:1990ux} serves as the bridge connecting the FV spectrum to the on-shell scattering K-matrix in the infinite volume (IFV):
\begin{equation}\label{eq:Lüscher}
{\rm det} [G_F^{-1}(L, E) - K(E)] = 0,
\end{equation}
where $G_F$ is  a known quantity that captures
 the finite volume kinematics and $K$ is the K-matrix in the infinite volume. 
However,  while this approach is generally model-independent, its validation requires a sufficiently large  box size $L$ to ensure that exponentially suppressed effects can be neglected. However, in practice, lattice simulations of systems with long-range interactions are often conducted in small boxes, causing discrepancies. One manifestation of this limitation is the left-hand cut (lhc) problem, where Lüscher’s quantization conditions break down in the presence of a nearby lhc. For instance, in systems like $\Lambda\Lambda$~\cite{Green:2021qol} and $DD^*$~\cite{Padmanath:2022cvl}, the lhc obstructs the use of certain energy levels for extracting scattering information. It has been estimated that a lattice box size exceeding 8 fm is necessary to mitigate lhc effects for the NN system at physical pion mass~\cite{Meng:2023bmz}. Several methods have been proposed to address these issues \cite{Meng:2021uhz, Meng:2023bmz, Meng:2024kkp,Raposo:2023oru, Raposo:2025dkb, Bubna:2024izx, Hansen:2024ffk, Dawid:2024oey}, but only the method combing chiral effective field theory (EFT) and the plane wave expansion in Refs.~\cite{Meng:2021uhz, Meng:2023bmz, Meng:2024kkp}, as illustrated in Fig.~\ref{fig:illu}, has so far been applied to fit real lattice QCD data. Recently, this method was also reformulated within the helicity basis~\cite{Yu:2025gzg}. In this work, we will focus on this method and its practical application to the $DD^*$ system.

 Another complication with Lüscher's formula arises from the breaking of rotational symmetry in cubic boxes, leading to energy levels that typically receive contributions from multiple partial waves. Due to partial wave mixing, there is no direct one-to-one correspondence between phase shifts and finite-volume energy levels. This necessitates a parameterization of the K-matrix to extract scattering information.
One common parameterization is the effective range expansion (ERE), but it is valid only within a narrow energy range constrained by the lhc \cite{Du:2023hlu, Du:2024snq}. The use of chiral EFT allows one to avoid the limitations of the ERE, while the implementation of the plane wave method, in addition, can effectively deal with partial wave mixing.

\begin{figure}[t]
\begin{center}
\includegraphics[width=0.7\textwidth]{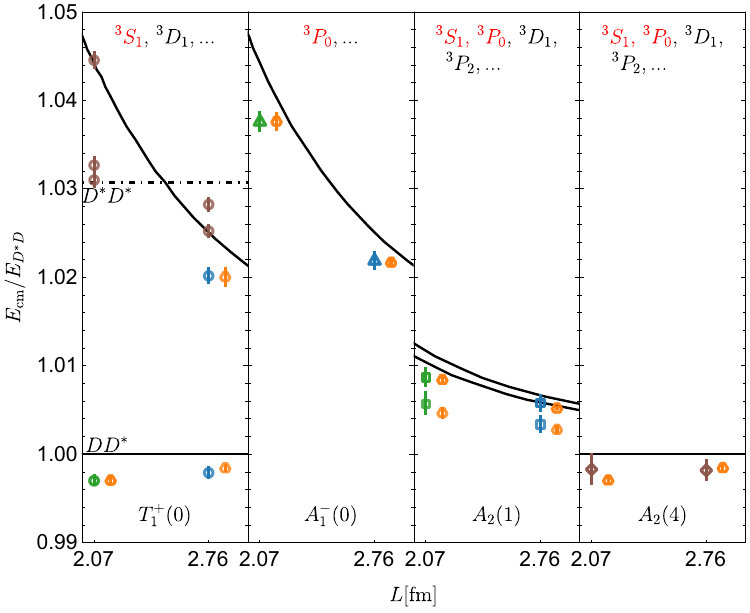}
\caption{ \label{fig:E_FV} Lattice data \cite{Padmanath:2022cvl} and fit results for the center-of-mass energy $ E_{\rm cm} = \sqrt{E^2 - {\bf P}^2} $ of the $ DD^* $ system, normalized by $ E_{DD^*} = M_D + M_{D^*} $, across various finite-volume irreps. Open circles, squares, and triangles represent the lattice energy levels, with blue and green points in the irreps $ T_1^+(0) $, $ A_1^-(0) $, and $ A_2(1) $ used as input. Orange symbols, slightly shifted to the right for clarity, represent the results of our full calculation (Fit 2), which includes pion effects. For each irrep, the contributing lowest partial waves are indicated. Predictions for the irrep $ A_2(4) $ are provided. Solid and dot-dashed lines denote the non-interacting $ DD^* $ and $ D^*D^* $ energies, respectively.}
\end{center}
\end{figure}

 $ DD^* $ scattering is used here as an example to demonstrate the power of our method. The lattice data is from Ref.~\cite{Padmanath:2022cvl}, where simulations were conducted with the lattice spacing of $ a \approx 0.08636 \, \text{fm} $ at $ m_{\pi} \approx 280 \, \text{MeV} $. The corresponding $ D $ and $ D^* $ meson masses are $ M_D = 1927 \, \text{MeV} $ and $ M_{D^*} = 2049 \, \text{MeV} $, respectively, with two spatial lattice sizes of $ L = 2.07 \, \text{fm} $ and $ L = 2.76 \, \text{fm} $, as shown in Fig.~\ref{fig:E_FV}. It is worth noting that the study of $ DD^* $ interactions is of significant importance. The LHCb collaboration reported the discovery of a promising $ DD^* $ molecular candidate, the $ T_{cc}(3875) $ state \cite{LHCb:2021auc, LHCb:2021vvq}. This $ T_{cc} $ state lies near the $ DD^* $ and $ DD\pi $ thresholds, featuring a minimal quark content of $ cc\bar{u}\bar{d} $. It serves as an excellent platform to explore ongoing topic in hadron physics, such as three-body dynamics, the left-hand cut problem, chiral dynamics, and the stability of tetraquark states, from various perspectives~\cite{Francis:2024fwf,Meng:2024kkp,Meng:2021jnw,Padmanath:2022cvl,Abolnikov:2024key,Lyu:2023xro,Du:2021zzh,Meng:2023jqk}.

\section{Left-hand cut problem}

\begin{figure}
    \centering
    \includegraphics[width=0.35\textwidth]{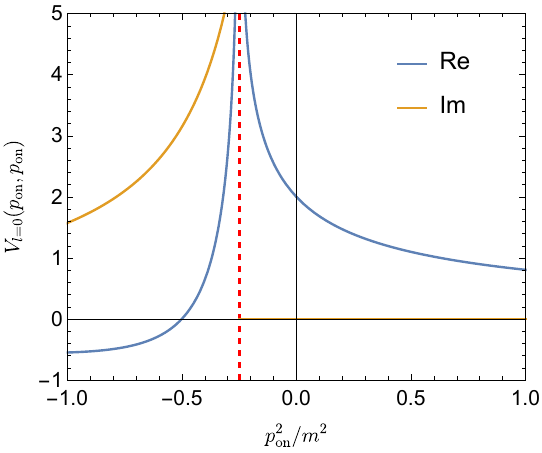}
    \caption{The S-wave on-shell potential, defined in Eq.~\eqref{eq:lhclog}. The red vertical dashed line marks the branch point of the left-hand cut at $ p_{\rm on}^{2}/m^2 = -\frac{1}{4} $. Below this point, both the potential and the corresponding $ K $-matrix become complex. The real and imaginary parts of the potential are depicted by blue and orange lines, respectively.  
    } 
    \label{fig:sm_Vlhc}
\end{figure}

To illustrate the left-hand cut problem, we closely follow Ref.~\cite{Meng:2023bmz}. We will use a long-range Yukawa-type interaction as an example. In momentum space, its form is given by  
\begin{equation}
V(\bm{p},\bm{p}') = \frac{1}{(\bm{p} - \bm{p}')^2 + m^2} = \frac{1}{p^2 + p'^2 - 2pp'z + m^2},\label{eq:yukawa}
\end{equation}
where $ m $ represents the mass of the exchanged particle, $\bm{p}$ and $\bm{p}'$ are the off-shell momenta, and $ z = \hat{p} \cdot \hat{p}' $. Performing a partial wave decomposition, for instance, for the S-wave potential, we obtain  
\begin{equation}
V_{l=0}(p, p') = \int_{-1}^1 \frac{dz}{p^2 + p'^2 - 2pp'z + m^2} = -\frac{1}{2pp'} \log\left(\frac{(p - p')^2 + m^2}{(p + p')^2 + m^2}\right).~\label{eq:lhclog}
\end{equation}
Evidently, the logarithmic function is multivalued due to the presence of branch cuts. We also introduce the on-shell momentum $ p_{\rm on} $, which, in the non-relativistic kinematics, is related to the energy by $ p_{\rm on}^2 = 2\mu_r E $, where $ \mu_r $ is the reduced mass of the two-body system.  
In the Lippmann-Schwinger equation (LSE), the 
momenta  ($p, p'$) are set to be positive or on-shell.  
Then, there is no singularity in the potential $ V(p, p') $ for $p, p' > 0$, nor in the half-on-shell potential $ V_{l=0}(p, p_{\rm on}) $ for $ p_{\rm on}^2 > 0 $. However, a singularity arises in the on-shell potential when $ p_{\rm on}^2 < -m^2/4 $. This can be derived by analyzing the denominator in the partial-wave decomposition of the on-shell potential (see Eq.~\eqref{eq:yukawa} with $p=p'=p_{\rm on}$):
\begin{equation}
2p_{\rm on}^2(1 - z) + m^2 = 0 \quad \Rightarrow \quad z = \frac{m^2}{2p_{\rm on}^2} + 1, \quad -1 < z < 1 \quad \Rightarrow \quad p_{\rm on}^2 < -\frac{m^2}{4}.
\end{equation}
This behavior is illustrated in Fig.~\ref{fig:sm_Vlhc}. For $ p_{\rm on}^2 < -m^2/4 $, an imaginary part appears. The same singularity also manifests itself in the on-shell $ K $-matrix, $ K_{l=0}(p_{\rm on}, p_{\rm on}) $.  In the infinite volume, the left-hand cut sets the convergence radius of the effective range expansion:
\begin{equation}
K^{-1}(p_{\rm on}) = p_{\rm on} \cot \delta(p_{\rm on}) = \frac{1}{a} + \frac{1}{2} r p_{\rm on}^2 + \cdots,\label{eq:ERE}
\end{equation}
as the analyticity assumption used in Eq.\eqref{eq:ERE} breaks down due to the emergence of the left-hand cut.  

The left-hand cut is a characteristic feature of long-range interactions. In the finite volume problem,  it leads to a breakdown of the Lüscher formula. Specifically, in the Lüscher condition, $ G_F^{-1}(L, E) $ is real. However, the presence of the left-hand cut introduces an imaginary component in $ K $ for $ p_{\rm on}^2 < -m^2/4 $, making Eq.~\eqref{eq:Lüscher} invalid. This demonstrates that the Lüscher formula cannot be naively applied to systems with long-range interactions. However, it is crucial to emphasize that, contrary to the on-shell potential, the half off-shell potential
$V_{l=0}(p_{\rm on},p')$  in Eq.~\eqref{eq:lhclog},  with real positive $p'$, as used in the LSE, is free of the pole in Eq.~\eqref{eq:yukawa} 
and the lhc, respectively,  when $E<0$. The same holds for the fully off-shell potential, ensuring that no issues arise when solving the Schrödinger equation as an eigenvalue problem in momentum space for $E<0$, which forms the foundation of our method. In addition, for the same reason, there is no lhc problem for the HAL QCD method~\cite{Aoki:2025jvi}.

\section{Formalism}

We begin with the Lippmann-Schwinger-type integral equations (LSE) in FV:
\begin{equation}
\mathbb{T}(E) = \mathbb{V}(E) + \mathbb{V}(E)\mathbb{G}(E)\mathbb{T}(E),
\end{equation}
where the FV energy levels are determined by solving the determinant equation:
\begin{equation}
\det\left[\mathbb{G}^{-1}(E) - \mathbb{V}(E)\right] = 0. \label{eq:rel_det0}
\end{equation}
The matrix in the determinant can be block-diagonalized according to the lattice irreducible representations (irreps). The discretized propagator $\mathbb{G}$ is defined as:
\begin{equation}
\mathbb{G}_{\bm{n},\bm{n}'} = {\cal J}\frac{1}{L^{3}}G(\tilde p_{\bm{n}},E)\delta_{\bm{n}',\bm{n}}, \quad G(\tilde p,E) = \frac{1}{4\omega_{1}\omega_{2}} \left( \frac{1}{E - \omega_1 - \omega_2} - \frac{1}{E + \omega_1 + \omega_2} \right),
\end{equation}
where $\mathcal{J}$ is the Jacobi determinant arising from the transformation between the box and center-of-mass frames, and $\tilde p_{\bm{n}}$ are the discretized momenta. The determinant equation can be reformulated as an eigenvalue problem, effectively solving the finite-volume Schrödinger-like equation. In this framework, the FV energy levels are interpreted as "bound states" confined within an effective potential well. 
%This method preserves the essential off-shell dynamics while circumventing the direct influence of the left-hand cut on the scattering parameters.

The interaction is constructed within chiral EFT (see Refs~\cite{Wang:2019ato,Wang:2019nvm,Wang:2020dko,Meng:2022ozq}. for general applications of the Chiral EFT), guided by chiral symmetry and its spontaneous breaking in QCD. Up to next-to-leading (NLO) order, the potential consists of short-range contact interactions (at LO and NLO) and the long-range LO one-pion exchange (OPE) interaction. The effective potential $V$ is expressed as:
\begin{equation}
V = V_{\text{OPE}}^{(0)} + V_{\text{cont}}^{(0)} + V_{\text{cont}}^{(2)} + \dots, \label{eq:veft}
\end{equation}
where the superscript denotes the power of $Q$ and $Q \sim m_{\pi}$ is the soft scale of the expansion. Two-pion exchange contributions at the considered $m_{\pi}$ value are assumed to be absorbed into the contact terms (see \cite{Chacko:2024ypl} for a related study). Truncating the contact interactions to $\mathcal{O}(Q^2)$, the relevant potentials contributing to irreps $T_1^+(0)$, $A_1^-(0)$, and $A_2(1)$ are:
\begin{equation}
\begin{aligned}
V_{\text{cont}}^{(0)+(2)}[^3S_1] &= \left(C^{(0)}_{^3S_1} + C^{(2)}_{^3S_1}(p^2 + p'^2)\right) (\bm{\epsilon} \cdot \bm{\epsilon}'^*), \\
V_{\text{cont}}^{(2)}[^3P_0] &= C^{(2)}_{^3P_0} (\bm{p}' \cdot \bm{\epsilon}'^*)(\bm{p} \cdot \bm{\epsilon}). \label{eq:Vct}
\end{aligned}
\end{equation}
Here, $\bm{p}$ ($\bm{p}'$) and $\bm{\epsilon}$ ($\bm{\epsilon}'$) represent the center-of-mass momenta and polarizations of the initial (final) $D^*$ meson, respectively. The longest-range interaction, driven by OPE, is given in the static approximation as:
\begin{equation}
V_{\text{OPE}}^{(0)} = -3\frac{M_D M_{D^*} g^2}{f_{\pi}^2} \frac{(\bm{k} \cdot \bm{\epsilon})(\bm{k} \cdot \bm{\epsilon}'^*)}{\bm{k}^2 + \mu^2}, \label{Eq:OPE}
\end{equation}
where $\mu^2 = m_{\pi}^2 - \Delta M^2$, $\Delta M = M_{D^*} - M_D$, and $\bm{k} = \bm{p}' + \bm{p}$. The pion decay constant $f_{\pi}$ is evaluated following Refs.~\cite{Du:2023hlu, Becirevic:2012pf}, yielding $f_{\pi} = 105.3$ MeV for $m_{\pi} = 280$ MeV. The coupling constant $g$ is extracted from a fit to physical point and the lattice data, giving $g = 0.517 \pm 0.015$ for the lattice spacing of $a \approx 0.086$ fm and $m_{\pi} = 280$ MeV. The lhc closest to the threshold has a branch point:
\begin{equation}
(p_{\lhc}^{1\pi})^2 = -\frac{\mu^2}{4} = -(126 \ \text{MeV})^2 \Rightarrow \left(\frac{p_{\lhc}^{1\pi}}{E_{DD^*}}\right)^2 \approx -0.001, \label{eq:lhc_num}
\end{equation}
where $E_{DD^*} = M_D + M_{D^*}$. While the OPE could induce a three-body right-hand cut corresponding to the $DD\pi$ state, for $m_{\pi} = 280$ MeV this contribution starts at $p_{\rhc_3}^2 = (552 \ \text{MeV})^2$, which is irrelevant for the current analysis. Notably, the OPE includes all partial waves in Eq.~\eqref{Eq:OPE}, as no partial wave expansion or truncation is applied in our plane wave expansion method.

The contact interactions in the LSE are regulated using exponential functions of the form $ e^{\frac{-(p^n+p'^n)}{\Lambda^n}} $ with $ n = 6 $. The regularization of operators involving the single pion propagator, which preserves long-range dynamics, follows the approach detailed in Ref.~\cite{Reinert:2017usi}. This is achieved through the substitution:  
\begin{equation}
\frac{1}{\bm{k}^2 + \mu^2} \to \frac{1}{\bm{k}^2 + \mu^2} e^{\frac{-(\bm{k}^2+\mu^2)}{\Lambda^2}}.
\end{equation}
In the following analysis, we use a cutoff $\Lambda = 0.9$ GeV. The associated uncertainties from the cutoff are small and are discussed in Ref.~\cite{Meng:2023bmz}. The unknown low-energy constants (LECs) are determined by fitting to lattice data at $ m_\pi = 280 $ MeV. This systematic expansion ensures a clear power-counting framework and manageable truncation errors.

\section{Numerical results}

To assess the impact of the OPE we perform two calculations. In Fit 1, we use a pure contact potential, adjusting the LECs $C^{(0)}_{^{3}S{1}}$, $C^{(2)}_{^{3}S{1}}$, and $C^{(2)}_{^{3}P{0}}$ to achieve the best $\chi^2$ fit to $E_{FV}$. In Fit 2, we include the OPE alongside contact interactions. The resulting energy levels for isoscalar $DD^*$ scattering are shown in Fig.\ref{fig:E_FV}. For both fits, partial-wave mixing is incorporated when calculating $E_{FV}$ in different lattice irreps. The OPE induces additional $S$-$D$ wave mixing due to long-range tensor interactions and introduces a new momentum scale associated with the lhc branch point in Eq.\eqref{eq:lhc_num}. This has significant effect on observables:
\begin{itemize}
    \item The OPE modifies the analytic structure of the scattering amplitude, rendering the phase shifts complex when analytically continued below the lhc;
    \item It dictates the energy dependence of the scattering amplitude near the threshold;
    \item It governs the leading exponentially suppressed corrections $\sim e^{-\mu L}$, which are neglected in the Lüscher approach.
\end{itemize}

\begin{figure}
    \centering
    \includegraphics[width=0.95\linewidth]{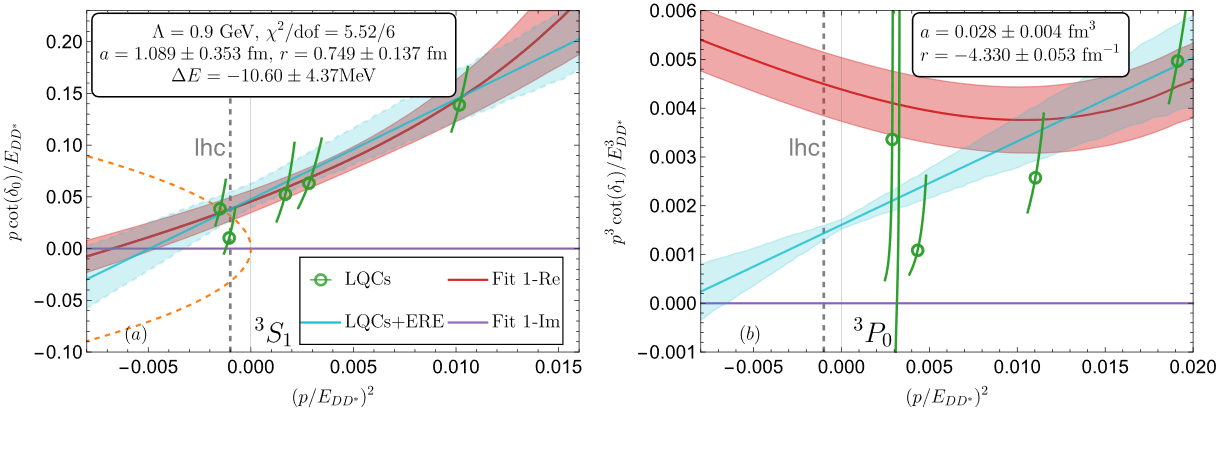}
      \includegraphics[width=0.95\linewidth]{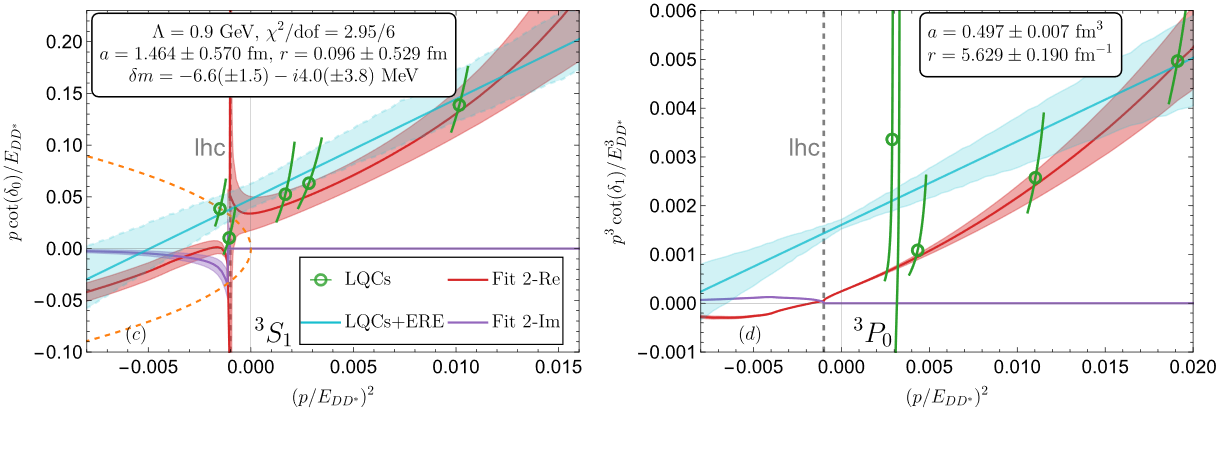}
    \caption{Phase shifts in the ${}^3S_1$ (leftpanel) and ${}^3P_0$ (right panel) partial waves extracted from lattice QCD data. Red bands represent the results of our 3-parameter fits without the OPE (Fit 1, upper  panel) and with the OPE (Fit 2, lower panel),  including the $1\sigma$ uncertainty. Green dots in the left panel are the phase shifts using the single-channel L\"uscher quantization conditions in Ref.~\cite{Padmanath:2022cvl}. Green dots in the right panel are extracted in this study using the same method. Blue bands are the results of the 4-parameter fits using the ERE in
Ref.~\cite{Padmanath:2022cvl}. Orange lines in the left panel correspond to $ip=\pm |p|$ from unitarity, normalized to $E_{DD^*}$.The gray vertical dashed line denotes the position of the branch point of the left-hand cut nearest to the threshold.}
    \label{fig:phase_shift}
\end{figure}

Figure~\ref{fig:phase_shift} presents the results for phase shifts, effective range parameters, and pole information in the $^3S_1$ and $^3P_0$ partial waves. The predictions of Fit 1 for $\delta_{^3 S_1}$ (upper left panel) align with the analysis of Ref.\cite{Padmanath:2022cvl} using the ERE (Eq.\eqref{eq:ERE}), yielding comparable ERE parameters and the $T_{cc}$ pole position. This similarity is expected, as both approaches use two parameters in this partial wave, matching the scattering length and effective range. In contrast, the contact fit results for $\delta_{^3 P_0}$ fail to describe all data points, as the low-energy phase shift behavior cannot be captured with a single-parameter fit. To address range corrections, Ref.\cite{Padmanath:2022cvl} introduced a two-parameter fit consistent with Eq.\eqref{eq:ERE}. However, this is unnecessary in our analysis since the OPE predominantly drives the range corrections in this channel, as shown in Fit 2 (lower right panel). The OPE also has a significant effect on $\delta_{^3 S_1}$, where the interplay between the repulsive OPE and attractive short-range interactions leads to a pole in $p \cot \delta_{^3 S_1}$ near the lhc, consistent with Ref.~\cite{Du:2023hlu}. This strongly impacts the ERE's validity range, the extracted scattering length and effective range, and the $T_{cc}$ pole position, which our calculation suggests is likely to correspond to a resonance state. Additionally, comparing the phase shifts extracted using the Lüscher method (green points) with Fit 2 reveals discrepancies, particularly for the two lowest-energy data points, which are heavily influenced by the lhc. Although above the $DD^*$ threshold, the Lüscher method agrees with our analysis for both $\delta_{^3 S_1}$ and $\delta_{^3 P_0}$ phase shifts within errors, the systematic uncertainties of the Lüscher method are difficult to quantify.

A similar EFT-based analysis of the isovector $DD^*$ scattering lattice data at $m_{\pi} = 280$ MeV is presented in Ref.~\cite{Meng:2024kkp}.

\section{Conclusion}

We present a novel approach based on EFT to extract two-body scattering information from finite-volume energies, utilizing the chiral expansion at low energies. Unlike the Lüscher method, our approach explicitly accounts for the longest-range interaction, by including the leading left-hand cut, thereby preserving the correct analytic structure of the scattering amplitude near the threshold. Finite-volume energy levels are directly calculated as eigenvalue solutions both below and above the left-hand cut. This method also addresses range effects and leading exponentially suppressed corrections in a model-independent manner.

Key computational advantages are achieved through the use of the plane wave basis, enabling efficient incorporation of partial-wave mixing effects on the lattice. The efficacy of our approach, as presented in \cite{Meng:2023bmz} and summarized here,  is demonstrated through a comprehensive analysis of lattice energy levels for $DD^*$ scattering~\cite{Padmanath:2022cvl}, relevant to the doubly charmed tetraquark. We find that the long-range interaction from the OPE significantly impacts the understanding of infinite-volume observables, 
%particularly in the $^3P_0$ channel, where 
as it enables proper amplitude calculation near the left-hand cut—a limitation of the Lüscher method. Systematic uncertainties due to truncation of the EFT expansion are shown in \cite{Meng:2023bmz} to be smaller than  statistical uncertainties. Our analysis indicates that the $T_{cc}^+$ state is likely a below-threshold resonance shifted into the complex plane due to OPE effects. With reduced uncertainties in the lattice energy levels, our approach can be used to extract the OPE strength, represented by $g/f_{\pi}$, directly from the lattice data. Furthermore, the incorporation of three-body ($DD\pi$) right-hand cuts is straightforward and will be critical for analyses at lower pion masses~\cite{Hansen:2024ffk}.

The novel method advocated in this contribution is broadly applicable to hadronic systems at unphysical pion masses, enhancing studies where finite-volume energy levels are available or forthcoming from lattice simulations. It can facilitate our understanding of nucleon-nucleon,  hyperon-nucleon and hyperon-hyperon scattering, nuclear structure, neutron stars, $\Sigma$ hypernuclei, D-mesic nuclei, and exotic hadrons such as tetraquarks, pentaquarks, and six-quark states.

\section*{Acknowledgment}
This work has been supported in part by the European Research Council (ERC AdG NuclearTheory, grant No.885150) and by the MKW NRW under the funding code NW21-024-A.

% \begin{thebibliography}{99}
% \bibitem{...}
% ....

% \end{thebibliography}

\bibliographystyle{JHEP}
\bibliography{ref.bib}

\end{document}